\documentclass{elsart}

\input epsf
\def\Dpi0pi0{$D^0 \rightarrow \pi^0\pi^0 \,$}
\def\Dpippi0{$D^+ \rightarrow \pi^+\pi^0 \,$}

\def\Dk0k0{$D^0 \rightarrow K^0\bar{K}^0 \,$}
\def\Dkpk0{$D^+ \rightarrow K^+\bar{K}^0 \,$}

\begin{document}

\begin{frontmatter}

\title{Study of the Cabibbo-suppressed decay modes
$D^0 \rightarrow \pi^-\pi^+$ and $D^0 \rightarrow K^-K^+ $}

%{\Large FOCUS AUTHOR LIST}
$\textrm{The~FOCUS~Collaboration}^\star$
\thanks{See \textrm{http://www-focus.fnal.gov/authors.html} for
additional author information.}
\author[ucd]{J.~M.~Link},
\author[ucd]{M.~Reyes},
\author[ucd]{P.~M.~Yager},
\author[cbpf]{J.~C.~Anjos},
\author[cbpf]{I.~Bediaga},
\author[cbpf]{C.~G\"obel},
\author[cbpf]{J.~Magnin},
\author[cbpf]{A.~Massafferri},
\author[cbpf]{J.~M.~de~Miranda},
\author[cbpf]{I.~M.~Pepe},
% ...
\address[ucd]{University of California, Davis, CA 95616} \address[cbpf]{Centro 
Brasileiro de Pesquisas F\'isicas, Rio de Janeiro, RJ, Brasil} \address[cinv]{CINVESTAV, 
07000 M\'exico City, DF, Mexico} \address[cu]{University of Colorado, Boulder, CO 80309}
\address[fnal]{Fermi National Accelerator Laboratory, Batavia, IL 60510} \address[fras]
{Laboratori Nazionali di Frascati dell'INFN, Frascati, Italy I-00044} \address[ui]
{University of Illinois, Urbana-Champaign, IL 61801} \address[iu]{Indiana University, 
Bloomington, IN 47405} \address[korea]{Korea University, Seoul, Korea 136-701} \address[kp]
{Kyungpook National University, Taegu, Korea 702-701} \address[milan]{INFN and University 
of Milano, Milano, Italy} \address[nc]{University of North Carolina, Asheville, NC 28804}
\address[pavia]{Dipartimento di Fisica Nucleare e Teorica and INFN, Pavia, Italy} \address[pr]
{University of Puerto Rico, Mayaguez, PR 00681} \address[sc]{University of South Carolina, 
Columbia, SC 29208} \address[ut]{University of Tennessee, Knoxville, TN 37996} \address[vu]
{Vanderbilt University, Nashville, TN 37235} \address[wisc]{University of Wisconsin, Madison, 
WI 53706}

\begin{abstract}
Using data from the FOCUS (E831) experiment at Fermilab, we present a new measurement
for the branching ratios of the Cabibbo-suppressed decay modes 
$D^0 \rightarrow \pi^-\pi^+$ and $D^0 \rightarrow K^-K^+ $. We measured:

$\Gamma(D^{0} \rightarrow K^-K^+)/\Gamma(D^{0}\rightarrow \pi^-\pi^+) = 2.81\pm 0.10\textrm{(stat)}
\pm 0.06\textrm{(syst)}$,
   
$\Gamma(D^{0} \rightarrow K^-K^+)/\Gamma(D^{0}\rightarrow K^-\pi^+) = 0.0993\pm 0.0014\textrm{(stat)}
\pm 0.0014\textrm{(syst)}$, and  

$\Gamma(D^{0} \rightarrow \pi^-\pi^+)/\Gamma(D^{0}\rightarrow K^-\pi^+)=0.0353\pm 0.0012\textrm{(stat)}
\pm 0.0006\textrm{(syst)}$.
 
 These values have been combined with other experimental data to extract the ratios of isospin
amplitudes and the phase shifts for the $D \rightarrow KK$ and $D \rightarrow \pi\pi$ decay 
channels.
\end{abstract}

\end{frontmatter}  
% PHG removed this

\textbf{1. Introduction}

In recent years, hadronic decays of charm mesons in many decay modes have
been extensively studied. The theoretical models that have been developed
mainly describe the 2-body decay modes and these models have led to several
successful predictions. However the branching ratio ${\frac{{\Gamma
(D^{0}\rightarrow K^{-}K^{+})}}{{\Gamma (D^{0}\rightarrow \pi ^{-}\pi ^{+})}}
}$ has, for a long time, been a puzzle of charm physics. The decay modes 
$D^{0}\rightarrow K^{-}K^{+}$ and $D^{0}\rightarrow \pi ^{-}\pi ^{+}$ are
both Cabibbo-suppressed and in first order perturbative calculation both 
receive contributions from the same diagrams (external spectator and exchange). 
To first order in the SU(3) flavour symmetry limit, the above
branching ratio should be one. However this ratio is reduced by a
factor $0.86$ due to a phase space difference and increased by a factor of 
$(f_{K}/f_{\pi })^{2}=1.49$ because of the different decay constants of the
kaon and the pion. An overall ratio of $1.29$ is thus expected. Including
SU(3) breaking effects, the expected ratio can increase to an upper limit of
about 1.4~\cite{BR_KKPP_OLD_1,BR_KKPP_OLD_2}.

However, the measured ratio is close to $2.5$~\cite{PDG}. Both penguin
diagrams and final state interactions (FSI) have been proposed as sources of
such a high value. However the penguin interference~\cite{PINGUINO} seems
to be unable to explain this large value. Some theoretical models~\cite{BR_KKPP_NEW_1,BR_KKPP_NEW_2} 
propose the FSI as the solution of this puzzle.

In this paper we present a new measurement of the ${\frac{{
\Gamma(D^{0} \rightarrow K^-K^+)} }{{\Gamma(D^{0} \rightarrow \pi^-\pi^+)}}}$
branching ratio obtained using data from the FOCUS experiment as well
as an isospin analysis of the $D \rightarrow KK$ and $D \rightarrow \pi\pi$
decay channels.

FOCUS is a charm photoproduction experiment~\cite{spectro} which collected 
data during the
1996--97 fixed target run at Fermilab. Electron and positron beams (with
typically $300~\textrm{GeV}$ endpoint energy) obtained from the $800~\textrm{GeV}$ Tevatron
proton beam produce, by means of bremsstrahlung, a photon beam which
interacts with a segmented BeO target. The mean photon energy for triggered
events is $\sim 180~\textrm{GeV}$. A system of three multicell threshold \v{C}erenkov
counters performs the charged particle identification, separating kaons from
pions up to $60~\textrm{GeV}/c$ of momentum. Two systems of silicon microvertex
detectors are used to track particles: the first system consists of 4 planes
of microstrips interleaved with the experimental target~\cite{WJohns} and the
second system consists of 12 planes of microstrips located downstream of the
target. These detectors provide high resolution in the transverse plane
(approximately $9~\mu\textrm{m}$), allowing the identification and separation of charm
primary (production) and secondary (decay) vertices. The charged particle
momentum is determined by measuring their deflections in two magnets of
opposite polarity through five stations of multiwire proportional chambers.

%%%%%%%%%%%%%%%%%%%%%%%%%%%%%%%%%%%%%%%%%%%%%%%%%%%%%%%%%%%%%%%%%%%%%%%%%%%%%%%%%%%
%

\vskip 0.5cm \textbf{2. Analysis of $D^0 \rightarrow \pi^-\pi^+$ and $D^0
\rightarrow K^-K^+$} %
%%%%%%%%%%%%%%%%%%%%%%%%%%%%%%%%%%%%%%%%%%%%%%%%%%%%%%%%%%%%%%%%%%%%%%%%%%%%%%%%%%%

The final states are selected using a \textit{candidate driven vertex
algorithm}~\cite{spectro}. A secondary vertex is formed from the two
candidate tracks. The momentum of the resultant $D^{0}$ candidate is used as
a \textit{seed} track to intersect the other reconstructed tracks and to 
search
for a primary vertex. The confidence levels of both vertices are
required to be greater than $1\%$. Two estimators of the relative isolation
of these vertices are returned by the algorithm: the first estimator (\emph{Iso1}) 
being the confidence level that tracks forming the secondary vertex
might come from the primary vertex, while the second estimator (\emph{Iso2})
is the confidence level that other tracks in the event might be associated
with the secondary vertex. Once the production and decay vertices are
determined, the distance $L$ between them and its error $\sigma _{L}$ are 
computed. The quantity $L$\thinspace /\thinspace $\sigma _{L}$ is an
unbiased measure of the significance of detachment between the primary and
secondary vertices. These variables provide a good measure of the
topological configuration of the event, so that appropriate cuts on them
reject the combinatorial background effectively.

In addition to large combinatorial backgrounds, $D^{0}\rightarrow \pi
^{-}\pi ^{+}$ and $D^{0}\rightarrow K^{-}K^{+}$ decays are difficult to
isolate because of a large reflection from the Cabibbo favored $
D^{0}\rightarrow K^{-}\pi ^{+}$ decays. Particle identification requirements
for each decay mode have been chosen to optimize signal quality. To minimize
systematic errors on the measurements of the branching ratios, we use
identical vertex cuts on the signal and normalizing modes. The only
difference in the selection criteria among different decay modes lies in the
particle identification cuts.

\begin{figure}[!!t]
%\vspace{15cm}
\epsfysize=18.cm \epsfxsize=8.cm \epsfbox{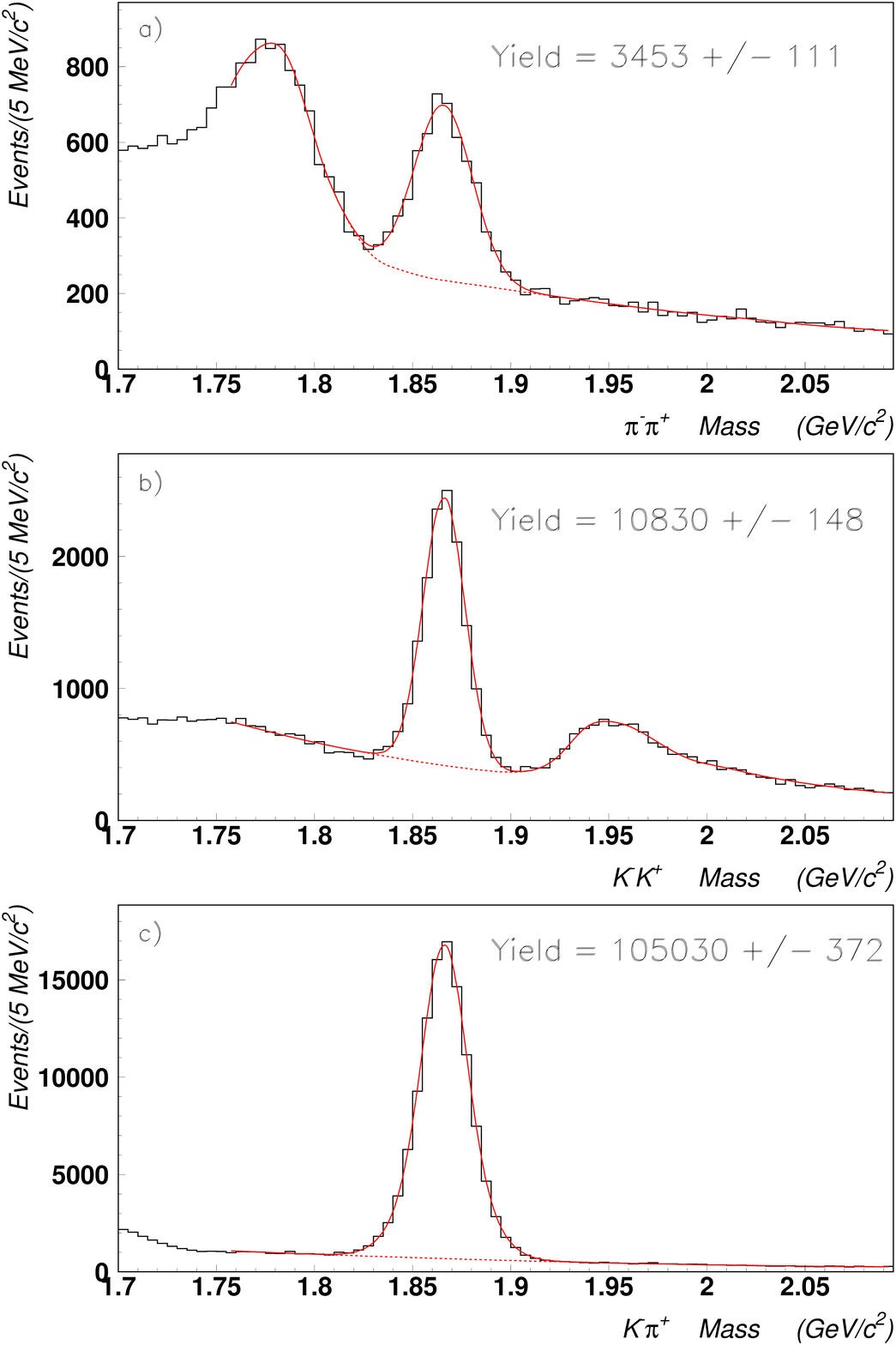} \vspace{0.5cm}
\caption{Invariant mass distribution for $\protect\pi^-\protect\pi^+$(a), $
K^-K^+$(b) and $K^-\protect\pi^+$(c). The fit (solid curve) for the
Cabibbo-suppressed decay modes of $D^0$ is to a gaussian over a polynomial
(for the combinatorial background) and a function obtained with Monte Carlo
simulations for the reflection peak.}
\label{2bodies}
\end{figure}

In the $D^{0}\rightarrow \pi ^{-}\pi ^{+}$ and $D^{0}\rightarrow K^{-}K^{+}$
analysis, we require $L$\thinspace /\thinspace $\sigma _{L}$ $>$ $10$, 
\emph{Iso1} $<10\%$ and \emph{Iso2} $<$ 0.5 $\%$. We also require the $D^{0}$
momentum to be in the range $25\rightarrow 250~\textrm{GeV}/c$ (a very loose cut) and
the primary vertex to be formed with at least two reconstructed tracks
in addition to the seed track. The \v{C}erenkov identification cuts used in
FOCUS are based on likelihood ratios between the various stable particle
identification hypotheses. These likelihoods are computed for a given track
from the observed firing response (on or off) of all the cells that are
within the track's ($\beta =1$) \v{C}erenkov cone for each of our three 
\v{C}erenkov counters. The product of all firing probabilities for all the cells
within the three \v{C}erenkov cones produces a $\chi ^{2}$-like variable 
$W_{i}=-2\ln (\mathrm{Likelihood})$ where $i$ ranges over the electron, pion,
kaon and proton hypotheses~\cite{cerenkov}. All the kaon tracks are required
to have $\Delta _{K}=W_{\pi }-W_{K}$ (kaonicity) greater than $1$, whereas all
the pion tracks are required to have $\Delta _{\pi }=W_{K}-W_{\pi }$ (pionicity)
exceeding $1.5$. Using the set of selection cuts just described, we obtain the
invariant mass distributions for $\pi ^{-}\pi ^{+}$, $K^{-}K^{+}$ and $K^{-}\pi ^{+}$
shown in Fig. \ref{2bodies}.

In Fig. 1a the $\pi ^{-}\pi ^{+}$ mass plot shows a broad peak to the left 
of the signal peak due to surviving contamination from $D^{0}\rightarrow
K^{-}\pi ^{+}$ events. The shape of the reflection peak has been determined by
generating Monte Carlo $D^{0}\rightarrow K^{-}\pi ^{+}$ events and
reconstructing them as $\pi ^{-}\pi ^{+}$. The mass plot is fit
with a function that includes a Gaussian for the signal, a third-order
polynomial for the combinatorial background and a shape for the reflection
background obtained from the Monte Carlo simulation. The amplitude of the
reflection peak is a fit parameter while its shape was fixed. The low mass
region is excluded in the fit to avoid possible contamination due to other
charm hadronic decays involving an additional $\pi^{0}$. A least squares
fit gives a signal of $3453\pm 111$$\pi ^{-}\pi ^{+}$ events.

The $K^{-}K^{+}$ mass plot, shown in Fig. 1b, is fit with a function
similar to that for the $\pi^-\pi^+$ fit. In the $K^{-}K^{+}$ case, the
reflection peak is on the right of the signal. The fit gives a signal of 
$10830 \pm 148$$K^{-}K^{+}$ events.

The large statistics $K^{-}\pi ^{+}$ mass plot of Fig. 1c is fit with two
Gaussians{\footnote{With the lower statistics of the $K^{-}K^{+}$ and 
$\pi ^{-}\pi ^{+}$ signals a single Gaussian gives an adequate fit.} 
with the same mean but different sigmas to take into account the different resolution
in momentum of our spectrometer~\cite{spectro} plus a second-order polynomial.
The fit gives a signal of $105030\pm 372$$K^{-}\pi ^{+}$ events. }

The fitted $D^0$ masses are in good agreement with the world average~\cite{PDG} 
and the widths are in good agreement with those of our Monte Carlo
simulation. 

%%%%%%%%%%%%%%%%%%%%%%%%%%%%%%%%%%%%%%%%%%%%%%%%%%%%%%%%%%%%%%%%%%%%%%%%%%%%%%%%
%
%                                  BRANCHING RATIOS
%
%%%%%%%%%%%%%%%%%%%%%%%%%%%%%%%%%%%%%%%%%%%%%%%%%%%%%%%%%%%%%%%%%%%%%%%%%%%%%%%%

\vskip 0.5cm \textbf{3. Relative Branching Ratios}

The evaluation of relative branching ratios requires yields from the fits to
be corrected for detection efficiencies, which differ among the various
decay modes because of differences in both spectrometer acceptance (due to
different $Q$ values for the various decay modes) and \v{C}erenkov
identification efficiency.

From the Monte Carlo simulations, we compute the relative efficiencies to 
be: 
${\frac{\epsilon(D^{0} \rightarrow \pi\pi) }{\epsilon(D^{0} \rightarrow KK)}}
= 0.897 \pm 0.003$, ${\frac{\epsilon(D^{0} \rightarrow K\pi) }{
\epsilon(D^{0} \rightarrow KK)}} = 0.963 \pm 0.003$ and ${\frac{
\epsilon(D^{0} \rightarrow K\pi) }{\epsilon(D^{0} \rightarrow \pi\pi)}} =
1.074 \pm 0.004$. Using the previous results, we obtain the following 
values
for the branching ratios: ${\frac{{\Gamma(D^{0} \rightarrow KK)} }{{
\Gamma(D^{0} \rightarrow \pi\pi)}}}  = 2.81 \pm 0.10$, ${\frac{{\Gamma(D^{0}
\rightarrow KK)} }{{\Gamma(D^{0} \rightarrow K\pi)}}}  = 0.0993 \pm 0.0014$, and 
${\frac{{\Gamma(D^{0} \rightarrow \pi\pi)} }{{\Gamma(D^{0} \rightarrow
K\pi)}}} = 0.0353 \pm 0.0012$.

Systematic uncertainties on branching ratio measurements can come from
different sources. We determine three independent contributions to the
systematic uncertainty: the \emph{split sample} component, the \emph{fit
variant} component, and the limited statistics of the Monte Carlo.

The \emph{split sample} component takes into account the systematics
introduced by a residual difference between data and Monte Carlo, due to a
possible mismatch in the reproduction of the $D^{0}$ momentum and the changing
experimental conditions of the spectrometer during data collection. This 
component has been determined by splitting data
into four independent subsamples, according to the $D^{0}$ momentum range
(high and low momentum) and the configuration of the vertex detector,
that is, before and after the insertion of an upstream silicon system. A technique,
employed in FOCUS and in the predecessor experiment E687, modeled after the 
\emph{S-factor method} from the Particle Data Group~\cite{PDG}, was used 
to try to separate true systematic variations from statistical 
fluctuations.
The branching ratio is evaluated for each of the $4~(=2^{2})$ statistically
independent subsamples and a \emph{scaled variance} $\tilde{\sigma}$ (that 
is, where the errors 
are boosted when $\chi ^{2}/(N-1)>1$) is calculated.  The \emph{split 
sample}
variance $\sigma_\textrm{split}$ is defined as the difference between the 
reported statistical
variance and the scaled variance, if the scaled variance exceeds the
statistical variance: 

\begin{eqnarray}
\sigma _\textrm{split} & = \sqrt{\tilde{\sigma}^{2}-\sigma _\textrm{stat}^{2}}\hspace{1cm}
& if\quad \tilde{\sigma}>\sigma _\textrm{stat} \\ 
\sigma _\textrm{split} & = 0\hspace{2.4cm}\quad & if\quad \tilde{\sigma}<\sigma _\textrm{stat} 
\quad . \nonumber
\end{eqnarray}

Another possible source of systematic uncertainty is the \emph{fit variant}.
This component is computed by varying, in a resonable manner, the fitting
conditions on the whole data set. In our study, we changed the background
parametrization (varying the degree of the polynomial), the fit function for
the reflection peak (the reflection shape from the Monte Carlo was replaced
by a Gaussian), and the use of two Gaussian for the fit of the peak of $
D^{0}\rightarrow \pi ^{-}\pi ^{+}$ and $D^{0}\rightarrow K^{-}K^{+}$. The
values obtained by the various fits are all a priori likely, therefore this
uncertainty can be estimated by the simple average of the measures of the fit 
variants: 
\begin{eqnarray}
\sigma _\textrm{fit}=\sqrt{\frac{\sum_{i=1}^{N}x_{i}^{2}-N \langle x \rangle^{2}}{N-1}}\quad .
\end{eqnarray}

Finally, there is a further contribution due to the limited statistics of 
the Monte Carlo simulation used to determine the efficiencies. Adding in
quadrature the three components, we get the final systematic 
errors summarized in Table \ref{err_sist_riass_2corpi}:

\begin{table}[htb!]
\begin{center}
\begin{tabular}{|l|c|c|}
\hline
{Source} & $\sigma_{\textrm{BR}(KK/K\pi)}$ & $\sigma_{\textrm{BR}(\pi\pi/K\pi)}$ \\ \hline
{Split sample} & $0.0005$ & $0.0000$ \\ 
{Fit variant} & $0.0012$ & $0.0006$ \\ 
{MC statistics} & $0.0003$ & $0.0001$ \\ \hline
{Total systematic } & $0.0014$ & $0.0006$ \\ \hline
\end{tabular}
\end{center}
\caption{Sources of uncertainty on the $\frac{\Gamma(K^-K^+)}{\Gamma(K^-
\protect\pi^+)}$ and $\frac{\Gamma(\protect\pi^-\protect\pi^+)}{\Gamma(K^-
\protect\pi^+)}$ branching ratios.}
\label{err_sist_riass_2corpi}
\end{table}

The final results are shown in Table \ref{BR_two body} along with a
comparison with the previous determinations.

\begin{table}[!h]
\begin{center}
\begin{tabular}{|l|l|l|l|}
\hline
Experiment & $\frac{\Gamma(D^0 \rightarrow K^-K^+)}{\Gamma(D^0 \rightarrow
K^-\pi^+)}$ & $\frac{\Gamma(D^0 \rightarrow \pi^-\pi^+)}{\Gamma(D^0
\rightarrow K^-\pi^+)}$ & $\frac{\Gamma(D^0 \rightarrow K^-K^+)}{\Gamma(D^0
\rightarrow \pi^-\pi^+)}$ \\ \hline
E687~\cite{E687} & $0.109 \pm 0.007 \pm 0.009$ & $0.043\pm 0.007 \pm 0.003$ & 
$2.53 \pm 0.46 \pm 0.19$ \\ 
E791~\cite{E791} & $0.109 \pm 0.003 \pm 0.003$ & $0.040 \pm 0.002 \pm 0.003$
& $2.75 \pm 0.15 \pm 0.16$ \\ 
CLEO~\cite{CLEO} & $0.1040 \pm 0.0033 \pm 0.0027$ & $0.0351 \pm 0.0016 \pm
0.0017$ & $2.96 \pm 0.16 \pm 0.15$ \\ 
E831 (this result) & $0.0993 \pm 0.0014 \pm 0.0014$ & $0.0353 \pm 0.0012 \pm
0.0006$ & $2.81 \pm 0.10 \pm 0.06$ \\ \hline
\end{tabular}
\end{center}
\caption{Comparison with other experiments.}
\label{BR_two body}
\end{table}

%%%%%%%%%%%%%%%%%%%%%%%%%%%%%%%%%%%%%%%%%%%%%%%%%%%%%%%%%%%%%%%%%%%%%%%%%%%%%%%%
%
%                                  ISOSPIN
%
%%%%%%%%%%%%%%%%%%%%%%%%%%%%%%%%%%%%%%%%%%%%%%%%%%%%%%%%%%%%%%%%%%%%%%%%%%%%%%%%

\vskip 0.5cm \textbf{4. Isospin analysis of $D \rightarrow K K$ and $D
\rightarrow \pi \pi$ channels}

Final State Interactions (FSI) can dramatically modify the observed decay
rates and complicate the comparison of the experimental data with the
theoretical predictions. By means of the isospin analysis of the decay
channels $D\rightarrow KK$ and $D\rightarrow \pi \pi $, we can gain some
insight on the elastic component of the FSI (pure rotation in isospin space).

Let us consider the $D \rightarrow \pi\pi$ transistions: $D^{0} \rightarrow
\pi^{-}\pi^{+}$, $D^{0} \rightarrow \pi^{0}\pi^{0}$ and $D^{+} \rightarrow
\pi^{+}\pi^{0}$. The decay amplitudes can be expressed in terms of isospin $
I_f=0~(A_0)$ and $I_f=2~(A_2)$ amplitudes. The final state with isospin $
I_f=1$ is forbidden by Bose statistics for an angular momentum zero system
of two pions. We denote by $A^{+-}$, $A^{00}$ and $A^{+0}$ the decay
amplitudes for the $D^0 \rightarrow \pi^-\pi^+ $, $D^0 \rightarrow
\pi^0\pi^0 $ and $D^+ \rightarrow \pi^+\pi^0 $, respectively. Expressing the
decay amplitude in terms of isospin amplitudes, 
we have~\cite{ISOSPINPP1,ISOSPINPP2,ISOSPINPP3}:

\begin{eqnarray}
A^{+-} & = & +\sqrt{\frac{2}{3}} |A_{0}|\exp{(i\delta_{0})} + \sqrt{\frac{1}{
3}} |A_{2}|\exp{(i\delta_{2})} \\
A^{00} & = & -\sqrt{\frac{1}{3}} |A_{0}|\exp{(i\delta_{0})} + \sqrt{\frac{2}{
3}} |A_{2}|\exp{(i\delta_{2})} \\
A^{+0} & = & +\sqrt{\frac{3}{2}} |A_{2}|\exp{(i\delta_{2})} \quad .
\end{eqnarray}

Adding the decay amplitudes in quadrature, we find the ratio of the
magnitude of isospin amplitudes and their relative phase shift difference in
terms of measured branching fractions:\footnote{The relationship between the 
isospin amplitude and the branching fraction is 
\hbox{$\Gamma^{ij}=\frac{1}{8 \pi} \frac{p^\star}{M_{D}^{2}}\mid
A^{ij}\mid^{2}$}, where $p^\star$ is the center of mass 3-momentum of each
final particle~\cite{PDG}.}

\begin{eqnarray}
\mid \frac {A_{2}}{ A_{0}} \mid ^{2} & = & \frac {\frac{2}{3}\mid A^{0+}\mid
^{2}} {\mid A^{+-}\mid^{2} + \mid A^{00}\mid^{2}- \frac{2}{3}\mid
A^{+0}\mid^{2}} \\
\cos(\delta_{2} -\delta_{0}) & = & \frac {3\mid A^{+-}\mid^{2} -6\mid
A^{00}\mid^{2} +2\mid A^{+0}\mid^{2}} {4\sqrt{2\mid A^{+0}\mid^{2}} \sqrt{
\frac{3}{2} (\mid A^{00}\mid^{2}+\mid A^{+-}\mid^{2}) - \mid A^{+0}\mid^{2}}}
\quad .
\end{eqnarray}

The decay rate $\Gamma (D^{0}\rightarrow \pi ^{-}\pi ^{+})$ has been
determined from our measurement of the branching ratio $\frac{\Gamma (\pi
^{-}\pi ^{+})}{\Gamma (K^{-}\pi ^{+})}$, whereas the other $D\rightarrow \pi
\pi $ decay rates and lifetimes have been taken from the Particle Data
Group compilation~\cite{PDG}.

The results are shown in Table \ref{ISOSPIN_table}. In contrast to $
K\rightarrow \pi \pi $ decays, where the transitions are dominated by the $
I=1/2$ amplitude ($\Delta~I=1/2$ rule), the $A_{2}$ amplitude in $
D\rightarrow \pi \pi $ is comparable to the $A_{0}$ amplitude. Furthermore,
there is a large phase shift difference between the isospin amplitudes.
According to Watson's theorem~\cite{WATSON}, this phase shift cannnot
arise from the weak processes alone and thus constitutes direct evidence for
FSI~\cite{Jim_Varenna}.

In the same way, we can consider the two-body $D \rightarrow KK$
transitions: $D^{0} \rightarrow K^{+}K^{-}$, $D^{0} \rightarrow K^{0}
\mbox{\={K}}^{0}$ and $D^{+} \rightarrow K^{+}\mbox{\={K}}^{0}$. The decay
amplitudes $A^{ij}$ for the $D \rightarrow K^iK^j$ decay modes can be
expressed in terms of $A_0$ and $A_1$ isospin amplitudes~\cite{ISOSPIN_KK}:

\begin{eqnarray}
A^{+-} & = & \frac{1}{\sqrt{2}} \left( |A_{1}|\exp{(i\delta_{1})} +
|A_{0}|\exp{(i\delta_{0})} \right) \\
A^{00} & = & \frac{1}{\sqrt{2}} \left( |A_{1}|\exp{(i\delta_{1})} -
|A_{0}|\exp{(i\delta_{0})} \right) \\
A^{+0} & = & \sqrt{2} |A_{1}|\exp{(i\delta_{1})}
\quad .
\end{eqnarray}

Using the previous decompositions, we can express the ratio of the 
magnitudes of the isospin amplitudes and their phase shift difference in 
terms of the measured branching fractions: 
\begin{eqnarray}
\mid \frac {A_{1}}{ A_{0}} \mid ^{2} & = & \frac {\mid A^{0+}\mid ^{2}} {
2\mid A^{+-}\mid^{2} +2\mid A^{00}\mid^{2} - \mid A^{+0}\mid^{2}} \\
\cos(\delta_{1} -\delta_{0}) & = & \frac {\mid A^{+-}\mid^{2} -\mid
A^{00}\mid^{2}} {\sqrt{\mid A^{+0}\mid^{2}} \sqrt{2\mid A^{00}\mid^{2}+2\mid
A^{+-}\mid^{2} - \mid A^{+0}\mid^{2}}}
\quad .
\end{eqnarray}

The $\Gamma (D^{0}\rightarrow K^{-}K^{+})$ decay rate has been determined
from our measurement of the branching ratio $\frac{\Gamma (K^{-}K^{+})}{
\Gamma (K^{-}\pi ^{+})}$, the $\Gamma (D^{+}\rightarrow K^{+}\mbox{\={K}}
^{0})$ from a previous measurement of FOCUS~\cite{BRIAN} and the remaining
decay rate from the Particle Data Group compilation~\cite{PDG}.

The results are shown in Table \ref{ISOSPIN_table}. Analogously to the  $D
\rightarrow \pi\pi$ case, the two $D \rightarrow K K$ isospin amplitudes 
are of the same order of magnitude, although the isospin phase shift 
difference is smaller.

The isospin analysis of the $D \rightarrow KK$ and $D \rightarrow \pi\pi$
decay channels is summarized in Table \ref{ISOSPIN_table} (the quoted
errors are obtained adding in quadrature the statistical and 
systematic errors) along with a comparison to previous determinations 
by CLEO~\cite{ISOSPINPP3,ISOSPIN_KK}:

\begin{table}[ht!]
\begin{center}
\begin{tabular}{|c|c|c|}
\hline
Quantity & CLEO & E831(this result) \\ 
\hline
$\mid A_2\mid /\mid A_0 \mid$ & $0.72 \pm 0.13 \pm 0.11$ & $0.65 \pm 0.14$
\\ 
$\delta_2-\delta_0$ & $(82.0 \pm 7.5 \pm 5.2)^{\circ}$ & $(83.6 \pm
10.0)^{\circ}$ \\ 
$\mid A_1\mid /\mid A_0 \mid$ & $0.61^{+0.11}_{-0.10}$ & $0.56 \pm 0.04$ \\ 
$\delta_1-\delta_0$ & $(28.4^{+12.1}_{-9.7})^{\circ}$ & $(37.1 \pm
7.5)^{\circ}$ \\ \hline
\end{tabular}
\end{center}
\caption{Isospin analysis for $D \rightarrow K K$ and $D \rightarrow \protect
\pi\protect\pi$ decay modes, where $\mid A_2\mid /\mid A_0 \mid$ and $
\protect\delta_2-\protect\delta_0$ refer to $D \rightarrow \protect\pi
\protect\pi$, while $\mid A_1\mid /\mid A_0 \mid$ and $\protect\delta_1-
\protect\delta_0$ to $D \rightarrow K K$.}
\label{ISOSPIN_table}
\end{table}

These results show that strong interactions, acting on the final particles,
play a very important role in $D\rightarrow KK$ and $D\rightarrow \pi \pi $
decays, modifying the measured $\frac{\Gamma (K^{-}K^{+})}{\Gamma (\pi
^{-}\pi ^{+})}$ ratio.

Another way to see the elastic FSI effect on the $\frac{\Gamma (K^{-}K^{+})}{
\Gamma (\pi ^{-}\pi ^{+})}$ branching ratio is to compute the ratio of the
sums over the $D^{0}$ isospin rotated decay modes~\cite{RAPPORTO}: $\frac{
\Gamma (K^{-}K^{+})+\Gamma (K^{0}\bar{K}^{0})}{\Gamma (\pi ^{-}\pi
^{+})+\Gamma (\pi ^{0}\pi ^{0})}$. As opposed to the $\frac{\Gamma
(K^{-}K^{+})}{\Gamma (\pi ^{-}\pi ^{+})}$ ratio, this ratio is not affected
by elastic FSI.

Using these measurements for $\frac{\Gamma(K^-K^+)}{\Gamma(K^-\pi^+)}$ and 
$
\frac{\Gamma(\pi^-\pi^+)}{\Gamma(K^-\pi^+)}$ and the PDG~\cite{PDG} values
for the other modes, we compute:

\begin{eqnarray}
\frac{\Gamma(K^{-}K^{+})+\Gamma(K^{0}\bar{K}^{0})} {\Gamma(\pi^{-}\pi^{+})+
\Gamma(\pi^{0}\pi^{0})} = 2.06 \pm 0.24
\quad .
\end{eqnarray}

This ratio is lower than the $\frac{\Gamma(K^-K^+)}{\Gamma(\pi^-\pi^+)}$
branching ratio, but still above the expected value of $1.4$. Therefore, 
the elastic FSI cannot account for all the discrepancy between theory and experiments.
An inelastic FSI that also allows the transition $KK \rightarrow \pi\pi$ seems
to be the most reasonable explanation~\cite{BR_KKPP_NEW_1}.

%%%%%%%%%%%%%%%%%%%%%%%%%%%%%%%%%%%%%%%%%%%%%%%%%%%%%%%%%%%%%%%%%%%%%%%%%%%%%%%%
%
%                                  CONCLUSIONS
%
%%%%%%%%%%%%%%%%%%%%%%%%%%%%%%%%%%%%%%%%%%%%%%%%%%%%%%%%%%%%%%%%%%%%%%%%%%%%%%%%

\vskip 1.5cm \textbf{4. Conclusions}

We have measured the following branching ratios: $\frac{\Gamma
(D^{0}\rightarrow K^{-}K^{+})}{\Gamma (D^{0}\rightarrow K^{-}\pi ^{+})}$, $
\frac{\Gamma (D^{0}\rightarrow \pi ^{-}\pi ^{+})}{\Gamma (D^{0}\rightarrow
K^{-}\pi ^{+})}$ and $\frac{\Gamma (D^{0}\rightarrow K^{-}K^{+})}{\Gamma
(D^{0}\rightarrow \pi ^{-}\pi ^{+})}$. A comparison with previous
determinations has been shown in Table \ref{BR_two body}. Our results
improve significantly the accuracy of these measurements.

An isospin analysis of the decay channels $D\rightarrow KK$ and $
D\rightarrow \pi \pi $ shows that final state interactions play an important 
role in these hadronic decay modes.

%%%%%%%%%%%%%%%%%%%%%%%%%%%%%%%%%%%%%%%%%%%%%%%%%%%%%%%%%%%%%%%%%%%%%%%%%%%%%%%
%
%                                  acknowledgements
%
%%%%%%%%%%%%%%%%%%%%%%%%%%%%%%%%%%%%%%%%%%%%%%%%%%%%%%%%%%%%%%%%%%%%%%%%%%%%%%%%

\vspace{1.cm}

We wish to acknowledge the assistance of the staffs of Fermi National
Accelerator Laboratory, the INFN of Italy, and the physics departments of
the collaborating institutions. This research was supported in part by the
U.~S. National Science Foundation, the U.~S. Department of Energy, the
Italian Istituto Nazionale di Fisica Nucleare and Ministero della Istruzione
Universit\`a e Ricerca, the Brazilian Conselho Nacional de Desenvolvimento
Cient\'{\i}fico e Tecnol\'ogico, CONACyT-M\'exico, and the Korea Research
Foundation of the Korean Ministry of Education.

%%%%%%%%%%%%%%%%%%%%%%%%%%%%%%%%%%%%%%%%%%%%%%%%%%%%%%%%%%%%%%%%%%%%%%%%%%%%%%%
%
%                                  REFERENCES
%
%%%%%%%%%%%%%%%%%%%%%%%%%%%%%%%%%%%%%%%%%%%%%%%%%%%%%%%%%%%%%%%%%%%%%%%%%%%%%%%

%%%%%%%%%%%%%%%%%%%%%%%%%%%%%%%%%%%%%%%%%%%%%%%%%%%%%%%%%%%%%%%%%%%%%%%%%%%%%%%
%
%                                  FIGURE
%
%%%%%%%%%%%%%%%%%%%%%%%%%%%%%%%%%%%%%%%%%%%%%%%%%%%%%%%%%%%%%%%%%%%%%%%%%%%%%%%

\newpage

\end{document}